\documentclass[runningheads]{llncs}
\usepackage{graphicx}
\usepackage{textcomp}
\usepackage{array}
\newcolumntype{C}[1]{>{\centering\let\newline\\\arraybackslash\hspace{0pt}}m{#1}}
\usepackage{xcolor}
\def\BibTeX{{\rm B\kern-.05em{\sc i\kern-.025em b}\kern-.08em
		T\kern-.1667em\lower.7ex\hbox{E}\kern-.125emX}}
\begin{document}
\title{Two Phase Authentication and VPN Based Secured Communication for IoT Home Networks}
%TWO PHASE AUTHENTICATION AND VPN BASED SECURED COMMUNICATION FOR IOT HOME NETWORKS
%
\titlerunning{Secured Communication for IoT Home Networks}
% If the paper title is too long for the running head, you can set
% an abbreviated paper title here
%
\author{Md Masuduzzaman\inst{1}\orcidID{0000-0001-8039-367} \and
Ashik Mahmud\inst{1}\orcidID{0000-0003-1358-2954} \and
Anik Islam\inst{2}\orcidID{0000-0002-6725-9805} \and
Md Mofijul Islam\inst{3}\orcidID{0000-0003-4207-5863}}
\authorrunning{Masuduzzaman et al.}
% First names are abbreviated in the running head.
% If there are more than two authors, 'et al.' is used.
%
\institute{American International University-Bangladesh, Dhaka, Bangladesh
\email{masud.prince@aiub.edu, mahmud.devops@gmail.com}\\ \and
Kumoh National Institute of Technology, Gumi 39177, South Korea
\email{anik.islam@kumoh.ac.kr}\\
 \and
University of Dhaka, Dhaka, Bangladesh\\
\email{akash.cse.du@gmail.com}}

\maketitle              % typeset the header of the contribution
\begin{abstract}
With the advancement of technology, devices, which are considered non-traditional in terms of internet capabilities, are now being embedded in microprocessors to communicate and these devices are known as IoT devices. This technology has enabled household devices to have the ability to communicate with the internet and a network comprising of such device can create a home IoT network. Such IoT devices are resource constrained and lack high-level security protocols. Thus, security becomes a major issue for such network systems. One way to secure the networks is through reliable authentication protocols and data transfer mechanism.  As the household devices are controllable by the users remotely, they are accessed over the internet. Therefore, there should also be a method to make the communication over the internet between IoT devices and the users more secured. This paper proposes a two-phase authentication protocol for authentication purposes and a VPN based secure channel creation for the communication of the devices in the network. Furthermore, the paper discusses the Elliptic Curve Cryptography as a viable alternative to RSA for a more efficient Key exchange mechanism for low-powered IoT devices in the network.


\end{abstract}

\section{Introduction} 
	\subsection{Background}
	The Internet of Things (IoT) is a system of interconnected devices, sensors and actuators etc. which work together in a network to reach a common goal~\cite{b1}. Such technology can be implemented in various ways to make our daily lives easier by placing internet capabilities in devices which are not regularly used as network devices. In recent years, the application of microprocessor-based controllers in devices ranging from toasters to airliners is being added to connect them to the internet~\cite{b4,b5}. With such advancements in IoT technology, it is showing potential to be deployed as consumer products for home usage~\cite{b5}. Providing internet capabilities to home devices, such as air conditioners, lights, fans, refrigerators etc., enables them to be controlled remotely. Such type of application can be called home IoT networks. As the devices connected to a home network and these devices can be controlled remotely which make these devices vulnerable to malicious attacks~\cite{b2,b4} and secure data transferring is also needed~\cite{b19}. So there need to add methods of authentication and security between the user and the devices in the network in order to prevent attacks~\cite{b5,b7}. Secure authentication of devices in a network includes a key exchange mechanism and handshake between the devices ~\cite{b3,b7,b12}. This can be done through centralized network system in which a central node is responsible for the security mechanisms~\cite{b7,b12} or through distributed networks where each node shares private keys and after successful handshake communication is established~\cite{b3,b7,b12}.
	\begin{figure}
			\centerline{\includegraphics[width=1\columnwidth]{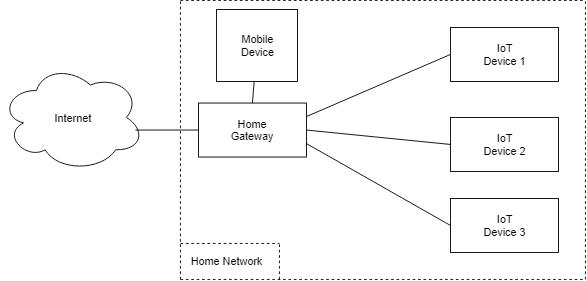}}
			\caption{System Step for Wi-Fi based security \cite{b5}.}
			\label{fig:1}
		\end{figure}

	 \subsection{Authentication Protocols}
	 Authenticating devices in a home network is a key process in securing the user interaction with IoT. If the network lacks security the end devices are vulnerable to attack and the purpose of implementing IoT is diminished~\cite{b13}.\\ There are several authentication mechanisms for IoT applications. Such as password based remote user authentication using one-way hash functions and ticket based authentication~\cite{b7}. Existing methods of authentication of devices include registering devices in a cloud platform running with the home network and initiating a handshake and key exchange~\cite{b7,b12}. Such systems also include a current method of key exchange which is the RSA key exchange mechanism~\cite{b6}. As most of the IoT devices are resource constrained, DTLS protocol is often used for handshaking~\cite{b16}.

	\begin{figure}
		\centerline{\includegraphics[width=\columnwidth]{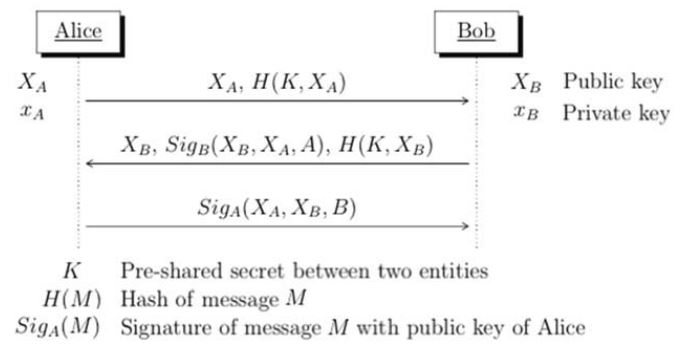}}
		\caption{Simple handshake protocol using ECC\cite{b5}.}
		\label{fig:2}
	\end{figure}
	\section{Related Works}
	
	\subsection{Wi-Fi Network Based Security}

	The system contains a home gateway, the user or mobile device and several IoT devices connected through a Wi-Fi network, as shown in Fig. \ref{fig:1}. Users can access the IoT devices from the home gateway and the gateway performs authentication and monitoring functions between the devices in the system~\cite{b5}. The authentication protocol used in this system involves the use of public key cryptography~\cite{b6} with pre-shared keys between the gateway and a new device which utilizes Elliptic Curve Cryptography (ECC) to reduce key size~\cite{b5}, as shown in Fig. \ref{fig:2}. This model lacks a proper mechanism for the handshaking protocol for constrained devices. Although ECC was used to reduce key size shared for authentication, the whole application lacks security measures if a malicious attacker is able to infiltrate the system.
	
	\subsection{PAuthKey Protocol}
\begin{figure}
		\centerline{\includegraphics[width=\columnwidth]{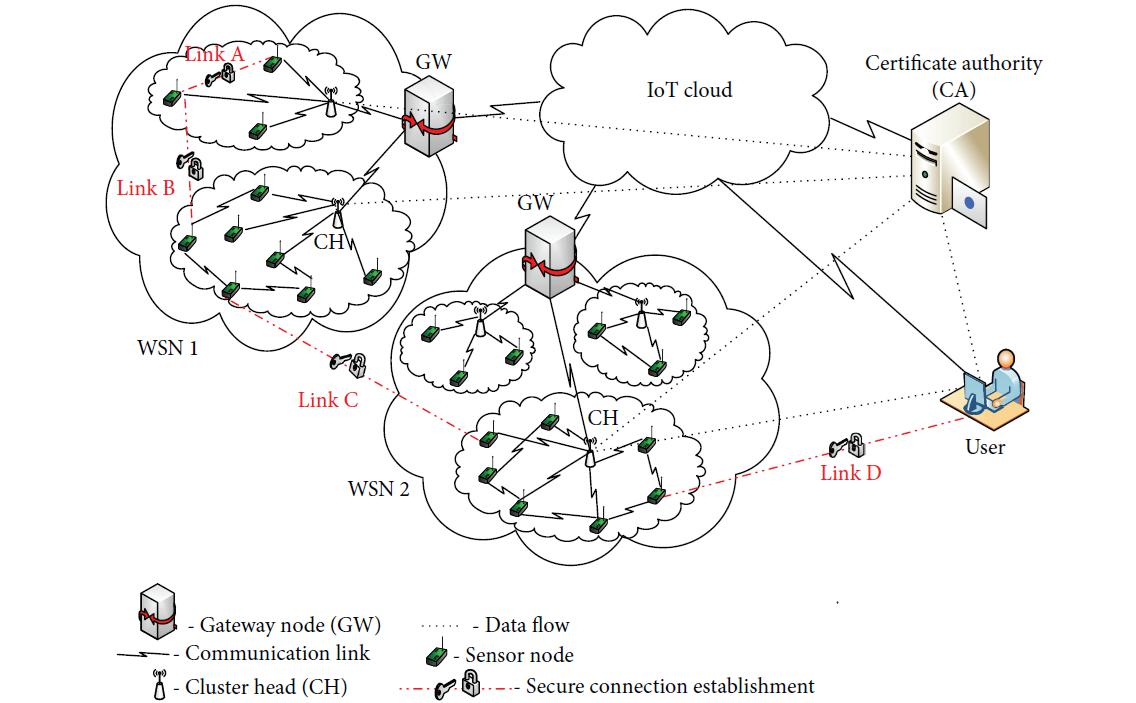}}
		\caption{PAuthKey Network Model\cite{b7}.}
		\label{fig:3}
		\end{figure}

	The system authenticates its devices in a two-phase authentication protocol~\cite{b7,b9,b12}, as shown in Fig. \ref{fig:3}. The network consists of several nodes in a cluster that communicates with the user over the IoT cloud through a gateway~\cite{b7}. A Certificate Authority (CA) is connected to the network and is responsible for authenticating the devices using handshake and key exchange~\cite{b7}.
	The first phase of the authentication is done manually by registering the devices in the certificate authority~\cite{b7,b12}. The second phase is done through initiating handshake using DTLS handshaking protocol and exchanging keys using ECC~\cite{b6,b7,b16}. Since the user communicates with the IoT devices over the IoT cloud or internet, in these areas, the data packets remain vulnerable to attackers. Even though they are encrypted, there still lie possibilities of the packets being sniffed and decrypted.
	
	\subsection{Two-way Authentication Security Scheme On Existing DTLS Protocol}
	
	\begin{figure}
		\centerline{\includegraphics[width=\columnwidth]{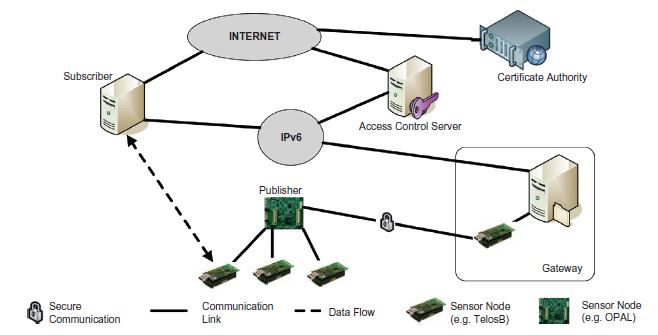}}
		\caption{DTLS based Security and Two-way Authentication Network Model\cite{b17}.}
		\label{fig:4}
	\end{figure}
	
\begin{figure}
		\centerline{\includegraphics[width=\columnwidth]{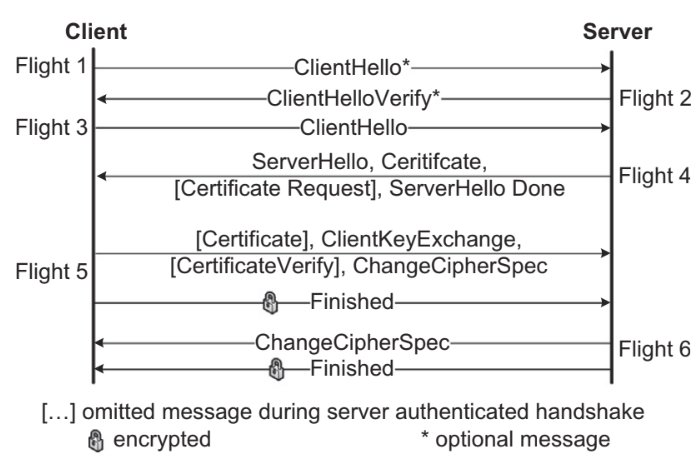}}
		\caption{DTLS Handshake Mechanism for Two-way Authentication\cite{b17}.}
		\label{fig:5}
	\end{figure}

	This system comprises of a network containing a certificate authority which is responsible for authorizing the devices and an access control server for exchanging key ~\cite{b17}, as shown in Fig. \ref{fig:4}. The devices communicate with each other over the internet and the IoT devices communicate with the user and the server through a gateway~\cite{b17}. The key aspect of this system is the use of DTLS handshake for mutual authentication of the devices, i.e. a two-way authentication~\cite{b9,b16,b17,b18}, as shown in Fig. \ref{fig:5}. During the handshake, keys are exchanged using RSA cryptography~\cite{b17}. This model also briefly talks about using VPN as a mechanism to secure payload over the internet for their proposed model~\cite{b13,b17,b20}. As this model uses the RSA key exchange mechanism, there exist the probability of large network overheads occurring due to large key size. This can be modified by using a key exchanging mechanism which generates a lower key size for the same security bit for RSA.

	\section{Network Model and Assumption}

	Our proposed network model follows the existing network model used in the PAuthKey system for the authentication process~\cite{b7}. Since the devices in their network are communicating over the internet or IoT cloud~\cite{b7,b11}, we modified the network using VPN for secure communication. Here, the model is made such that the wireless sensor networks (WSN) clusters with gateways (GW) can be an individual house or a room so that the system is scalable to larger implementations. The certificate authority used in PAuthKey system~\cite{b7} has been replaced by a VPN server which is responsible for registering the devices as well as authenticating the devices using DTLS handshake~\cite{b7,b17} and public key cryptography for key exchange mechanism~\cite{b6}. The VPN server also establishes VPN endpoint to gateway tunnels~\cite{b21,b22} for securing data packets sent by any devices in the system. The VPN server is responsible for creating VPN tunnels for data communication between the user and the device or the gateways and the server. The VPN server is responsible for creating VPN tunnels for data communication between the user and the device or the gateways and the server.
		
	\section{Proposed Solution}
	
	\subsection{MAC Address Based User Registration}
\begin{figure}
		\centerline{\includegraphics[width=1\columnwidth]{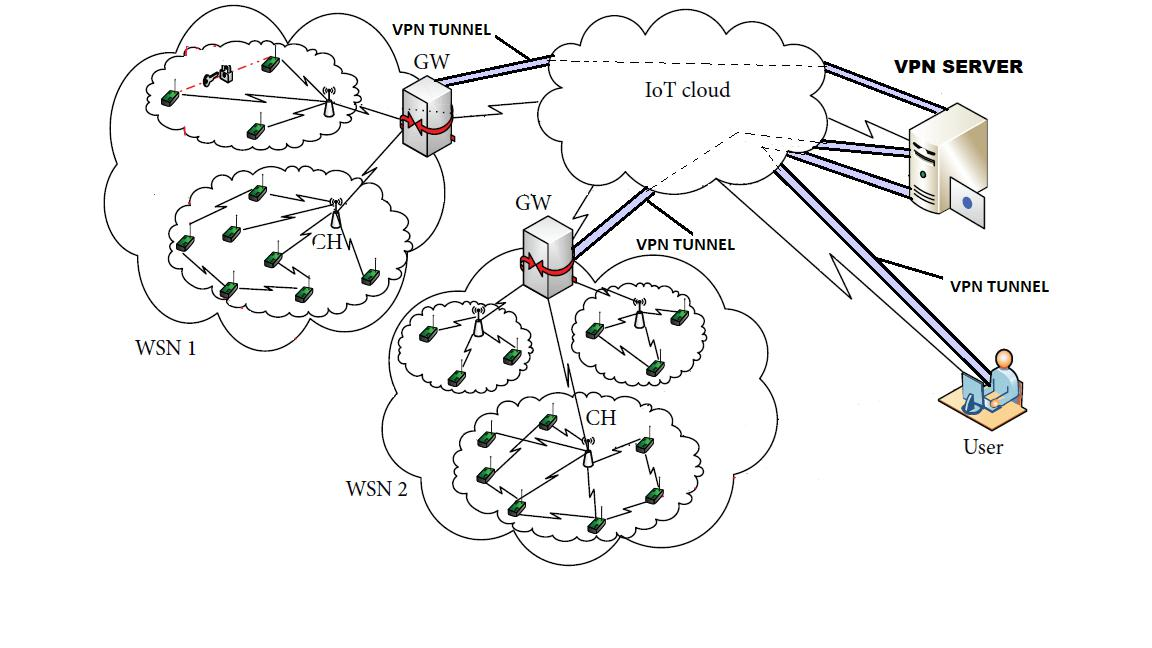}}
		\caption{Proposed Network Model.}
		\label{fig:6}
	\end{figure}

	Similar to the aforementioned systems, our system also uses a registration phase for the devices in the network~\cite{b5,b7,b17}, as shown in Fig. \ref{fig:6}. In our system, the VPN server contains a database that keeps a record of valid users which includes their usernames, password and MAC address, as shown in Table \ref{long}. This record is kept as a fail-safe such that if an authorized user does manage to gain access to the system, the server can deny them access as their MAC address does not match with any registered addresses.
	\subsection{Authentication Protocol}
	Use of ECC is now being adopted as an alternate public key exchange mechanism for IoT based networks~\cite{b5,b15}. In PAuthKey~\cite{b7} and Wi-Fi based system~\cite{b5}, the usage of ECC was also adopted along with DTLS handshake. On the other hand, the DTLS two-way authentication protocol used RSA as their method for key exchange. In our proposed method, we use ECC along with DTLS handshake for authentication purposes. The reason ECC is used in lieu of RSA is that ECC is more suitable for resource-constrained devices. ECC keys are generated through computation on an elliptical curve whose basic equation is
	\begin{equation}
	y^2=x^2+ax+b \label{eq}
	\end{equation}
	The trapdoor function of ECC is similar to RSA with complex mathematical computation [14], this is due to elliptic curves having horizontal symmetry and any non-vertical line will intersect the curve in 3 places at most.	Moreover, the endpoint of the non-vertical intersection can be reflected to form another non-vertical intersection resulting in the scope of more key generation~\cite{b14,b15}. The major advantages that ECC has over RSA are: 

\begin{itemize} 
\item ECC relies on difficult discrete logarithm functions which make it more difficult to decrypt by malicious attackers~\cite{b14}. 
\item ECC generates keys with shorter size compared to RSA for the same security bits ~\cite{b14,b15}.

\item For the same bit size, ECC generates more number of keys than RSA~\cite{b15}, which is provided in Table \ref{Table:1}.

 \end{itemize}
		
		Considering these advantages our proposed solution follows the ECC key exchange mechanism.  
		\begin{table}
			\caption{Comparable key sizes between RSA and ECC~\cite{b15}. }
				\label{Table:1}
			\begin{center}
				
				\begin{tabular}{|c|c|c|}
					
					\hline
					Security bits  & ECC & RSA\\
					\hline 
					80 & 260 & 1024\\
					\hline
					112	& 224 & 2048\\
					\hline
					128 & 256 & 3072\\
					\hline
					192 & 384 & 7680\\
					\hline
					256 & 521 & 15350\\
					\hline
					
					% Please Enter your data in the above code.
				\end{tabular}
			\end{center}
			\end{table}
		
		\subsection{Data Transfer Security Using VPN}
		In our network model and the network models discussed in our related works, users communicate with IoT devices through the internet or IoT cloud~\cite{b5,b7,b17}. This may leave the data packets being sent vulnerable to eavesdropping or sniffing attacks. To solve this problem, our model introduces VPN to the network. A VPN endpoint to gateway tunnel~\cite{b20} which is created from the server to the gateways, from user to server and from user to gateway by the VPN server as shown in Fig. 4. This enables the devices to communicate using private IP’s over a secured channel. Furthermore, such type of secured communication ensures that CIA (confidentiality, Integrity, Authenticity) is maintained in the system.

		\section{Conclusions}

	\subsection{Scope of Future Work}
		
		This solution focuses on secure data communication and authentication protocols for home IoT networks. The proposed network model mainly focuses on home networks including control over multiple homes for one user based on cluster topology. This network is scalable to larger network areas such as industries and hospitals and even smart cities. Furthermore, the scope of improving VPN based communication over the internet is vast as this solution focus on conceptual based discussion rather than practical implementation. 
		
		\subsection{Conclusion} 
		
		In this solution, the problem that was focused on the authentication mechanism of devices in a home IoT network and secured communication of devices over the internet. Efficient authenticating mechanism based on ECC and DTLS handshake was introduced and VPN based tunneling was proposed to secure data communication. Moreover, MAC address-based fail-safe solution was also proposed such that in an unlikely event an unauthorized user gains access to the system, they can be dealt with.
		
		\section*{Acknowledgement}

This research was supported by the Department of Science \& Information Technology, American International University of Bangladesh (AIUB). The authors are grateful for this support.

\end{document}